\documentclass[12pt]{article}
\usepackage{amssymb,amsmath,amscd}
\usepackage{}

\begin{document}

\topmargin -2pt


\headheight 0pt

\topskip 0mm \addtolength{\baselineskip}{0.20\baselineskip}
\begin{flushright}
{\tt KIAS-P07073}
\end{flushright}

\vspace{5mm}

\begin{center}
{\Large \bf Charged Rotating Black Holes on DGP Brane} \\

\vspace{10mm}

{\sc Daeho Lee}${}^{  \dag,  }$\footnote{dhlee@sju.ac.kr}, {\sc Ee Chang-Young}${}^{ \dag,
 }$\footnote{cylee@sejong.ac.kr}, and {\sc Myungseok Yoon}${}^{  \ddag,  }$\footnote{younms@sogang.ac.kr}
\\

\vspace{1mm}

 ${}^{\dag}${\it Department of Physics, Sejong University, Seoul 143-747, Korea}\\

${}^{\ddag}${\it Center for Quantum Spacetime, Sogang
    University, Seoul 121-742, Korea}\\

\vspace{10mm}

{\bf ABSTRACT} \\
\end{center}


\noindent

We consider charged rotating black holes localized on a three-brane
in the DGP model. Assuming a $Z_2$-symmetry across the brane and
with a stationary and axisymmetric metric ansatz on the brane,
a particular solution is obtained in the Kerr-Schild form.
This solution belongs to the accelerated branch of the DGP model
and has the characteristic of the Kerr-Newman-de Sitter type solution
in general relativity.
Using a modified version of Boyer-Lindquist coordinates we
examine the structures of the horizon and ergosphere.
\\

\vfill


\thispagestyle{empty}

\newpage




%

\section{Introduction}

Recent astronomical observations indicate that our universe is in
the phase of accelerated expansion \cite{sauv}. There has been
much recent interest in the idea that our universe may be a brane
embedded in some higher dimensional space. One of the models along this line
proposed by Dvali, Gabadadze and Porrati (DGP)
\cite{dgp} is known to contain a branch of solutions exhibiting
self accelerated expansion of the universe \cite{df}.

The brane world black holes in the Randall-Sundrum(RS) model
\cite{RS12} have been studied by many authors. Firstly,  Chamblin
{\it et al.} \cite{chr:prd} presented evidence that a non-rotating
uncharged black hole on the brane is described by a ``black
cigar'' solution in five dimensions. Then, Dadhich {\it et al.}
\cite{dmpr:plb} showed that the Reissner-N\"{o}rdstrom metric is
an exact solution of the effective Einstein equations on the
brane, and Shiromizu {\it et al.} \cite{sms:prd} derived the
effective gravitational equations on the brane.
A solution for charged brane world black holes in the RS model was
obtained in \cite{crss:prd} and the charged rotating case was
obtained in \cite{ag:prd}. In particular, Aliev {\it et al.}
\cite{ag:prd} found exact solutions of charged rotating black
holes in the Kerr-Schild form \cite{ks} using a stationary and
axisymmetric metric ansatz on the brane.

In the case of the DGP model, approximate Schwarzschild solutions had been
obtained in \cite{po,ls,ms,nr}. An exact Schwarzschild solution
on the brane was obtained in \cite{gi:prd}. In the DGP model, the sources
are assumed to be localized on a brane by a certain mechanism not related
to gravity itself.
In \cite{kol:prd}, it was discussed that if the sources were not localized,
the brane with the induced graviton kinetic term has effectively repulsive
gravity and it would push any source off the brane. As a result, ordinary
black holes cannot be held on the brane. However, authors of \cite{gi:prd}
commented that charged black holes could still be quasilocalized if the
corresponding gauge fields are localized.
 Recently, motivated by the above suggestion,  an exact
solution of charged black holes on the brane in the DGP model was obtained in \cite{el}.
However, up to now no solution of charged rotating black holes on the
 DGP brane is obtained.

Here, we intend to improve this situation a bit. We try to obtain
an exact solution of charged rotating black holes on the  brane in
the DGP model by noting a particular set of conditions that
satisfies  the constraint equation. We first obtain the solution
in the Kerr-Schild form \cite{ks}. Then, by using a modified
Boyer-Lindquist coordinate transformation we find the horizon and
ergosphere. In doing this, we use a stationary and axisymmetric
metric ansatz on the brane and the solution exhibits the
characteristics of self accelerated expansion of the brane world
universe.

This paper is organized as follows. In section 2, we set the
action and equations of motion of the DGP model following the
approach of Ref. \cite{ag:cqg}. In section 3, we get a rotating black hole solution
 on the brane in the absence of Maxwell field and
examine the properties of the solution. In section 4,
we extend the result of section 3 and find a solution
for the charged rotating case.
In section 5, we conclude with discussion.
In this last section, we discuss a possible bulk solution
consistent with our on-brane solution.
\\

\section{Action and field equations}

The DGP gravitational action in the presence of sources takes the
form \cite{dgp}
\begin{eqnarray}
\label{action}
S = M_{*}^{3}\int d^5 x \sqrt{-g}~^{(5)}R + \int d^4 x \sqrt{-h}
\left(M_{P}^{2}R+L_{matter}\right),
\end{eqnarray}
where $R$ and $^{(5)}R$ are the 4D and 5D Ricci scalars,
respectively and $L_{matter}$ is the Lagrangian of the matter
fields trapped on the brane. Here, the $(4+1)$ coordinates are
$x^{A}=(x^{\mu},y(= x^5))$, $\mu=0,1,2,3$, and $g$ is the determinant
of the five-dimensional metric $g_{AB}$, while $h$ is the determinant
of the four-dimensional metric $h_{\mu\nu}=g_{\mu\nu}(x^{\mu},y=0)$.
A cross-over scale is defined by $r_{c}=
m_{c}^{-1}=M_{P}^{2}/2M_{*}^{3}$.
There is a boundary(a brane) at $y=0$ and $Z_{2}$ symmetry across
the boundary is assumed.
 The field equations derived from the action (\ref{action}) have
the form
\begin{equation}
\label{eom:munu}
^{(5)}G_{AB}= ^{(5)}R_{AB}-\frac{1}{2}g_{AB}~^{(5)}R=
\kappa_{5}^{2}\sqrt{\frac{h}{g}}\left(X_{AB}+T_{AB}\right)
\delta(y),
\end{equation}
where $\kappa_{4}^{2}=M_{P}^{-2}$ and $\kappa_{5}^{2}=M_{*}^{-3}$,
while $
X_{AB}=-\delta_{A}^{\mu}\delta_{B}^{\nu}G_{\mu\nu}/\kappa_{4}^{2}$
and $T_{AB}= \delta_{A}^{\mu}\delta_{B}^{\nu}T_{\mu\nu}$ is the
energy-momentum tensor in the braneworld.

 Now, we consider the metric of the following form \cite{ag:cqg,dgl:jcap},
\begin{equation}
\label{metric}
ds^2=g_{AB}dx^{A}dx^{B}=g_{\mu\nu}(x,y)dx^{\mu}dx^{\nu}+2N_{\mu}
dx^{\mu}dy+(N^2+g_{\mu\nu}N^{\mu}N^{\nu})dy^2.
\end{equation}

The $(\mu 5)$, $(55)$ components of the field equations
(\ref{eom:munu}) are called as the momentum and Hamiltonian
constraint equations, respectively, and are given by
\cite{ag:cqg,dgl:jcap}
\begin{equation}
\label{eom:5i}
\nabla_{\nu}K^{\nu}_{~\mu}-\nabla_{\mu}K=0,
\end{equation}
\begin{equation}
\label{eom:55}
R-K^{2}+K_{\mu\nu}K^{\mu\nu}=0,
\end{equation}
where $K_{\mu\nu}$ is the extrinsic curvature tensor defined
by
\begin{equation}
\label{kmunu}
K_{\mu\nu}=\frac{1}{2N}(\partial_{y} g_{\mu\nu}-\nabla_{\mu}
N_{\nu}-\nabla_{\nu}N_{\mu}),
\end{equation}
and $\nabla_{\mu}$ is the covariant derivative operator
associated with the metric $g_{\mu\nu}$.

Integrating both sides of the field equation (\ref{eom:munu})
along the $y$ direction and taking the limit of $y=0$ on the
both sides of the brane we arrive at the Israel's junction
condition \cite{israel} on the $Z_2$ symmetric brane in the
relation \cite{el}

\begin{equation}
\label{jc}
G_{\mu\nu}=\kappa_{4}^2 T_{\mu\nu}+m_{c}(K_{\mu\nu}
-h_{\mu\nu}K).
\end{equation}

In this paper, we take the electro-magnetic field as the
matter source on the brane.
Using (\ref{jc}) in the constraint (\ref{eom:5i}) and
(\ref{eom:55}) we find that the momentum constraint equation
is satisfied identically, while the Hamiltonian constraint
equation is written as
\begin{eqnarray}
\label{Hamiltonc}
 R_{\mu\nu}R^{\mu\nu}-\frac{1}{3}R^{2}+ m_c^2 R+
 \kappa_{4}^{4}T_{\mu\nu}T^{\mu\nu}
 -2\kappa_{4}^2 R_{\mu\nu}T^{\mu\nu}=0,
\end{eqnarray}
where we used $T=T^{\mu}_{~\mu}=0$.

Finally, combining the Einstein Equations in the
bulk($y\neq 0$)
\begin{equation}
\label{5deinstein}
^{(5)}G_{AB}= ^{(5)}R_{AB}-\frac{1}{2}g_{AB}~^{(5)}R=0
\end{equation}
with (\ref{jc}) we arrive at the gravitational field
equations on the brane \cite{el}

\begin{eqnarray}
\label{grave}
G_{\mu\nu} &=&-E_{\mu\nu}-\frac{\kappa_{4}^{4}}{m_{c}^2}
(T^{\rho}_{~\mu}T_{\rho\nu}-\frac{1}{2}h_{\mu\nu}
T_{\rho\sigma}T^{\rho\sigma})
      \nonumber \\
     && -\frac{1}{m_{c}^2}(R^{\rho}_{~\mu}R_{\rho\nu}
-\frac{2}{3}RR_{\mu\nu}+\frac{1}{4}h_{\mu\nu}R^2
-\frac{1}{2}h_{\mu\nu}R_{\rho\sigma}R^{\rho\sigma})
  \nonumber \\
 &&+\frac{\kappa_{4}^2}{m_{c}^2}(R^{\rho}_{~\mu}T_{\rho\nu}
+T^{\rho}_{~\mu}R_{\rho\nu}-\frac{2}{3}RT_{\mu\nu}
-h_{\mu\nu}R_{\rho\sigma}T^{\rho\sigma}),
\end{eqnarray}
where $E_{\mu\nu}$ is the traceless ``electric part'' of the
5-dimensional Weyl tensor $^{(5)}\!C_{ABCD}$ \cite{sms:prd}
and $m_{c}^{-1}=\kappa_{5}^2 /2\kappa_{4}^2$.
In what follows we shall set $\kappa_{4}^2=8\pi$.

In general, the field equations on the brane are not closed
and one needs to solve the evolution equations into the bulk.
However, by assuming a special ansatz for the induced metric
on the brane, one can make the system of equations on the
brane closed.

\section{Rotating black hole solution}

We start with a stationary and axisymmetric metric describing a
rotating black hole localized on a 3-brane in the DGP model. We
write it as the Kerr-Schild form \cite{ks} in which the metric is
expressed in a linear approximation around the flat metric:
\begin{eqnarray}
\label{ks}
ds^2 =(ds^2)_{\texttt{flat}}+f(k_{\mu}dx^{\mu})^2,
\end{eqnarray}
where $f$ is an arbitrary scalar function and $k_{\mu}$ is a
null, geodesic vector field in both the flat and full metrics
with
\begin{equation}
k_{\mu}k^{\mu}=0, k^{\nu}D_{\nu}k_{\mu}=0.
\end{equation}
 Introducing the Kerr-Schild coordinates $x^{\mu}=\{u,r,\theta,
\varphi\}$, we write the metric  as \cite{ag:prd}
\begin{eqnarray}
\label{metricsol}
ds^2 &=& h_{\mu\nu}dx^{\mu}dx^{\nu}= [-(du+dr)^2+dr^2+
\Sigma d\theta^2  \nonumber \\
&& +(r^2+a^2)\sin^{2}\theta d\varphi^2+2a\sin^2 \theta
drd\varphi] +H(r,\theta)(du-a\sin^2\theta d\varphi)^2,
\end{eqnarray}
where
\begin{equation}
\label{sigma}
\Sigma(r,\theta)=r^2+a^2\cos^2\theta ,
\end{equation}
and $a$ is the angular momentum per unit mass of the black
hole.

For the uncharged case we can set $T_{\mu\nu}=0$, and the Hamiltonian constraint equation
(\ref{Hamiltonc}) is reduced to
\begin{equation}
\label{HamiltonT=0}
R_{\mu\nu}R^{\mu\nu}-\frac{1}{3}R^2+m_{c}^2 R=0.
\end{equation}
Note that the above equation is satisfied with the following two sets of conditions,
\begin{equation}
\label{flat}
R=0,~~R_{\mu\nu}R^{\mu\nu}=0,
\end{equation}
and
\begin{equation}
\label{ds}
R=12m_c^2,~~R_{\mu\nu}R^{\mu\nu}=36m_c^4.
\end{equation}
The first set (\ref{flat}) is satisfied with the metric function $H(r, \theta)=2Mr/\Sigma$
in the metric (\ref{metricsol}), which  is the usual Kerr solution in general relativity.
The second set (\ref{ds}) corresponds to a non-flat(de-Sitter) case,
in which the conditions (\ref{ds}) in terms of the metric function $H(r,\theta)$ in (\ref{metricsol})
are given by the following:
\begin{eqnarray}
\label{eq:rel1}
&& 12m_{c}^2 = \frac{\partial^2 H}{\partial r^2}
+\frac{4r}{\Sigma}\frac{\partial H}{\partial r}
+\frac{2H}{\Sigma},
\end{eqnarray}
\begin{eqnarray}
\label{eq:rel2}
36m_{c}^4 &=& \frac{4H}{\Sigma^2}\left(r\frac{\partial H}
{\partial r}+a^2\cos^2\theta\frac{\partial^2 H}
{\partial r^2}\right)+\frac{4r^2}{\Sigma^2}
\left(\frac{\partial H}{\partial r}\right)^2
+\frac{2r}{\Sigma}\frac{\partial H}{\partial r}
\frac{\partial^2 H}{\partial r^2} \nonumber \\
&& +\frac{1}{2}\left(\frac{\partial^2 H}{\partial r^2}
\right)^2+\frac{2}{\Sigma^4}(r^4-2a^2r^2\cos^2\theta
+5a^4\cos^4\theta)H^2.
\end{eqnarray}
The metric function $H$ satisfying the above two equations is
given by
\begin{equation}
\label{scalarH}
H=\frac{2Mr+m_{c}^2(r^4+6r^2a^2\cos^{2}\theta-3a^4 \cos^{4}
\theta)}{\Sigma},
\end{equation}
where the parameter $M$ is an arbitrary constant of integration.
One can easily check that the metric (\ref{metricsol}) with
(\ref{scalarH}) satisfies the equation (\ref{Hamiltonc})
with $T_{\mu\nu}=0$.
In the limit $a \rightarrow 0$, the metric (\ref{metricsol}) with
(\ref{scalarH}) is reduced to the Schwarzschild-de Sitter black
hole solution with the cosmological constant $\Lambda=3m_{c}^2$ in
general relativity.
This corresponds to the solution for the $U(r)=-2$ case in Ref. \cite{gi:prd},
and belongs to the accelerated
branch of Kerr-de Sitter type solution \cite{gi:plb,dehghani:prd}.

In order to check the physical properties of the metric given by (\ref{metricsol})
with (\ref{scalarH}) we want to transform the Kerr-Schild form to
the Boyer-Lindquist coordinates. However, the equations (\ref{eq:rel1})
and (\ref{eq:rel2}) are not preserved under the transformation for the usual Boyer-Lindquist
coordinates.
Thus in order to
preserve the equations (\ref{eq:rel1}) and (\ref{eq:rel2}) under coordinate transformation,
we use the following modified transformation of Boyer-Lindquist type:
\begin{equation}
\label{toBL} du=dt-\frac{r^2+a^2}{\Delta}dr- X d\theta,~
d\varphi=d\phi-\frac{a}{\Delta}dr -Y d\theta,
\end{equation}
where $\Delta=r^2+a^2-H(r,\theta)\Sigma(r,\theta)$, $X$ and $Y$ are
functions of $r$ and $\theta$ only and satisfy the relation
\begin{equation}
\label{XandY} \frac{\partial }{\partial r}X(r,\theta)
=\frac{\partial}{\partial \theta}
\left(\frac{r^2+a^2}{\Delta}\right), ~ \frac{\partial } {\partial
r}Y(r,\theta) =\frac{\partial}{\partial \theta}
\left(\frac{a}{\Delta}\right).
\end{equation}
The newly added terms $X$ and $Y$ can be analytically integrated from
the transformation (\ref{toBL}) for the function $H$ given in (\ref{scalarH}).
In fact, the modified transformation (\ref{toBL}) that satisfy (\ref{XandY})
leaves the Hamiltonian constraint (\ref{HamiltonT=0}) invariant.
Under the transformation (\ref{toBL}), the metric (\ref{metricsol}) takes
the form
\begin{eqnarray}
\label{BLT=0}
ds^2 &=& -(1-H)dt^2 +\frac{\Sigma}{\Delta} dr^2
         +2[X-H(X-Y a\sin^2\theta)]dtd\theta
         \nonumber \\
      &&+[\Sigma-(1-H)X^2 -2HXY a\sin^2\theta+Y^2(r^2+a^2+Ha^2\sin^2
      \theta)]d\theta^2
       \nonumber \\
      &&-2[(r^2+a^2)Y -a H(X-Y a\sin^2\theta)]\sin^2\theta d\theta
      d\phi +(r^2+a^2+H a^2 \sin^2\theta)\sin^2\theta d\phi^2
       \nonumber \\
     && -2Ha\sin^2\theta dt d\phi.
\end{eqnarray}
Using ``MATHEMATICA'' we check that the equations
(\ref{eq:rel1}) and (\ref{eq:rel2}) remain unchanged with the metric
(\ref{BLT=0}) for any metric function
$H(r,\theta)$. Hence, we can use (\ref{scalarH}) as a solution for
the metric in (\ref{BLT=0}).

For $r \ll r_c$, $X$ and $Y$ with (\ref{scalarH}) can be approximately written as
\begin{eqnarray}
X &\approx & -6 m_c^2a^2\sin2\theta \left[r-\frac{(r_1^2 +a^2)(r_1^2 -a^2\cos^2\theta )}{(r-r_1)(r_1 -r_2)^2}+\frac{2T_1\ln(\frac{r}{r_1}-1)}{(r_1 -r_2)^3} + (r_1 \leftrightarrow r_2)\right]+\eta_1(\theta), \nonumber \\
Y &\approx& 6 m_c^2a^3\sin2\theta \left[\frac{(r_1^2 -a^2\cos^2\theta )}{(r-r_1)(r_1 -r_2)^2}+\frac{2T_2\ln(\frac{r}{r_1}-1)}{(r_1 -r_2)^3} + (r_1 \leftrightarrow r_2)\right]+\eta_2(\theta),
\end{eqnarray}
where $T_1 =r_1^4-2r_1^3r_2-r_1r_2a^2\sin^2\theta+a^4\cos^2\theta$, $T_2=r_1r_2 -a^2\cos^2\theta$ and $r_1$, $r_2$ are two roots of the equation $r^2-2Mr+a^2=0$.
When the crossover scale $r_{c}$ is infinite (or $m_{c}=0$) and both $\eta_1(\theta)$ and $\eta_2(\theta) $ are set to be zero, then
 the equations (\ref{eq:rel1}) and (\ref{eq:rel2}) are preserved with the metric function $ H=\frac{2Mr}{\Sigma} $.
This corresponds to the exact Kerr solution in general relativity once we identify the pararmeter $M$ as the mass of the black hole.

The governing equation for horizon radius is given by
\begin{equation}
\label{horizon:con}
\Delta=r^2+a^2-2Mr-m_{c}^2(r^4+6r^2a^2\cos^{2}\theta-3a^4 \cos^{4}
\theta)=0.
\end{equation}
The metric (\ref{BLT=0}) with (\ref{scalarH}) has three horizons located
at $r_{\pm}$ and $r_{CH}$, provided the total mass $M$ lies in the range
$M_{1e}|_{\theta=\pi/2} \leq M \leq M_{2e}|_{\theta=0,\pi}$ where
$M_{1e}$ and $M_{2e}$ are given by
\begin{eqnarray}
\label{extrememass}
M_{1e}=\frac{1}{3\sqrt{6}m_{c}} \sqrt{\alpha-A^{3/2}},
~~ M_{2e}=\frac{1}{3\sqrt{6}m_{c}} \sqrt{\alpha+A^{3/2}}
\end{eqnarray}
with
\begin{equation}
\alpha=1+36m_{c}^2 a^2 -18m_{c}^2 a^2 \cos^2 \theta(1+12m_{c}^2 a^2)
+216m_{c}^4 a^4\cos^4 \theta(1-4m_{c}^2a^2\cos^2 \theta)
\end{equation}
and
\begin{equation}
A=1-12m_{c}^2a^2 (1+\cos^2\theta).
\end{equation}
Here $r_{CH}$, which is smaller than the crossover scale $r_{c}$,
is a cosmological horizon, $r_{+}$ and $r_{-}$ are outer and inner
horizon, respectively.

The horizons can be expressed explicitly as follows:
\begin{equation}
\label{horizon:zeroq} r_{\pm}=\frac{1}{2m_c}\left(D^{1/2}_{-} \pm
\sqrt{D_{+} -4Mm_{c}D^{-1/2}_{-} }\right), ~
r_{CH}=\frac{1}{2m_c}\left(-D^{1/2}_{-} + \sqrt{D_{+}
+4Mm_{c}D^{-1/2}_{-} }\right),
\end{equation}
where
\begin{equation}
D_{\pm}=C \pm \frac{1}{3}
\left[ C+A \left(\frac{2}{B+\sqrt{B^2-4A^3}}\right)^{1/3}
+\left(\frac{2}{B+\sqrt{B^2-4A^3}}\right)^{-1/3}\right]
\end{equation}
with $C=1-6m^2_c a^2\cos^2\theta$ and $B=2C[C^2+36m^2_ca^2
(1+3m^2_c a^2\cos^4\theta)]-108m^2_c M^2$.
Note that the horizons $r_{\pm}$ and $r_{CH}$ always have real
positive values if the total mass lies between the masses
$M_{1e}|_{\theta=\pi/2}$ and $M_{2e}|_{\theta=0,\pi}$.

For $r \ll r_c$, the outer and inner horizons can be approximated as
\begin{equation}
r_{\pm} \approx \frac{M \pm \sqrt{M^2-a^2(1-3m_c^2a^2\cos^2\theta)}}{1-6m_c^2 a^2 \cos^2\theta}.
\end{equation}
In the limit of $r_{c} \rightarrow \infty$, we get $M_{1e}
\rightarrow a$, $M_{2e} \rightarrow \infty$, $r_{\pm} \rightarrow M
\pm \sqrt{M^2-a^2}$, and $r_{CH} \rightarrow \infty$ independent
of the angle $\theta$ as one can expect from (\ref{horizon:con}).

 For the ergosphere we calculate the condition
\begin{equation}
\label{ergo}
g_{tt}=r^2+a^2\cos^2\theta-2Mr-m_c^2(r^4+6r^2a^2\cos^{2}\theta-3a^4 \cos^{4}
\theta) = 0,
\end{equation}
which has three boundaries of the ergosphere known as the ``static limit''
surfaces located at $r_E^{\pm}$ and $r_E^{CH}$.

 Setting $r_{E}^{\pm}$ and $r_{E}^{CH}$ as
\begin{equation}
r_E^{\pm}=\frac{1}{2m_c}\left(\tilde{D}^{1/2}_{-} \pm
\sqrt{\tilde{D}_{+}-4Mm_{c}\tilde{D}^{-1/2}_{-} }\right),~
r_E^{CH}=\frac{1}{2m_c}\left(-\tilde{D}^{1/2}_{-}
+ \sqrt{\tilde{D}_{+}+4Mm_{c}\tilde{D}^{-1/2}_{-} }\right),
\end{equation}
where
\begin{equation}
\tilde{D}_{\pm}=C \pm \frac{1}{3} \left[ C+\tilde{A} \left(\frac{2}
{\tilde{B}+\sqrt{\tilde{B}^2-4\tilde{A}^3}}\right)^{1/3}
+\left(\frac{2}{\tilde{B}+\sqrt{\tilde{B}^2-4\tilde{A}^3}}
\right)^{-1/3}\right]
\end{equation}
with $\tilde{A}=1-24m_c^2 a^2\cos^2\theta$ and
$\tilde{B}=2C[C^2+36m_c^2 a^2\cos^2\theta(1+3m_c^2
a^2\cos^2\theta)] -108m^2_c M^2$,
 we obtain
\begin{equation}
\label{inequal}
r_E^{-} \leq r_{-} < r_{+} \leq r_E^{+} < r_E^{CH} \leq r_{CH},
\end{equation}
 with the equalities holding at $\theta=0,\pi$.

Comparing this with the event horizons (\ref{horizon:zeroq}),
we see that the ergosphere lies in the region
$r_{+}< r< r_E^{+}$ and $r_E^{CH} <r<r_{CH}$, which coincide with the horizon
at $\theta=0,\pi$.
\\

\section{Charged rotating black hole solution}

In this section, we consider the case in which the brane contains a
Maxwell field with an electric charge. Outside the black hole,
the Maxwell field can be described by a source-free Maxwell equations.
Thus, we have to solve simultaneously the constraint equation (\ref{Hamiltonc}) and the Maxwell equations:
\begin{equation}
\label{fMaxwell}
g^{\mu\nu}D_{\mu}F_{\nu\sigma}=0,
\end{equation}
\begin{equation}
\label{sMaxwell} D_{[\mu}F_{\nu\sigma]}=0,
\end{equation}
where $D_{\mu}$ is the covariant derivative operator associated
with the brane metric $h_{\mu\nu}$.
However, we only need to
solve Eqs. (\ref{Hamiltonc}) and (\ref{fMaxwell}), since Eq. (\ref{sMaxwell}) is satisfied identically.

 Hinted from the characteristic of the Kerr-Newman solution in general relativity which yields $R_{\mu\nu}R^{\mu\nu}=4Q^4/\Sigma^4, R_{\mu\nu}T^{\mu\nu}=Q^4/2\pi\Sigma^4, T_{\mu\nu}T^{\mu\nu}=Q^4/16\pi^2\Sigma^4$,
we note that the Hamiltonian constraint (\ref{Hamiltonc}) on the brane is satisfied with the following two set of conditions:
\begin{equation}
\label{flatq}
R=0,~~ R_{\mu\nu}R^{\mu\nu}=\frac{4Q^4}{\Sigma^4}, ~~
R_{\mu\nu}T^{\mu\nu}=\frac{Q^4}{2\pi\Sigma^4},~~
T_{\mu\nu}T^{\mu\nu}=\frac{Q^4}{16\pi^2\Sigma^4}
\end{equation}
and
\begin{equation}
\label{dsq}
R=12m_c^2,~~ R_{\mu\nu}R^{\mu\nu}=36m_c^4+ \frac{4Q^4}
{\Sigma^4},
R_{\mu\nu}T^{\mu\nu}=\frac{Q^4}{2\pi\Sigma^4},~~
T_{\mu\nu}T^{\mu\nu}=\frac{Q^4}{16\pi^2\Sigma^4}.
\end{equation}
The first set (\ref{flatq}) is satisfied with the conventional Kerr-Newman solution which
is given by the following
potential one-form $A_{\mu}$ and the metric function $H$:
\begin{equation}
\label{poten}
A_{\mu}dx^{\mu}=-\frac{Q r}{\Sigma}(du-a\sin^2\theta d\varphi),
\end{equation}
\begin{equation}
\label{Hwithcharge}
H=\frac{2Mr-Q^2}{\Sigma},
\end{equation}
where the parameter $Q$ is the electric charge of the black hole.
The second set (\ref{dsq}) is satisfied with the following solution:
\begin{equation}
\label{potential}
A_{\mu}dx^{\mu}=-\frac{Q r}{\Sigma}(du-a\sin^2\theta d\varphi),
\end{equation}
\begin{equation}
\label{Hwithcharge}
H=\frac{2Mr-Q^2+m_{c}^2(r^4+6r^2a^2\cos^{2}\theta-3a^4 \cos^{4}
\theta)}{\Sigma}.
\end{equation}
In the non-rotating limit, $a\rightarrow0$, the second solution, (\ref{potential})
and (\ref{Hwithcharge}), reduces to the conventional charged de Sitter solution
with the cosmological constant $\Lambda=3m_{c}^2$ \cite{stu:prd}.
This case corresponds to the
$U(r)=-2$ case of Ref. \cite{el}, in which it was shown to belong to the accelerated branch.

In order to check the physical properties of the solution, we
again make a transformation of the Boyer-Lindquist type (\ref{toBL}).
Since the Maxwell equation (\ref{fMaxwell}) should transform
covariantly under (\ref{toBL}), the potential one-form
(\ref{potential}) should also transform covariantly:
\begin{equation}
\label{potential:trans} A_{\mu}dx^{\mu}= -\frac{Q
r}{\Sigma}\left[dt-a\sin^2\theta d\phi -(X-Y
a\sin^2\theta)d\theta\right]+\frac{Q r}{\Delta} dr.
\end{equation}
The nonvanishing components of the electromagnetic field tensor
$F_{\mu\nu}$ are given by
\begin{eqnarray}
\label{maxwell:field}
F_{r\theta}&=&-\frac{Q(r^2-a^2\cos^2\theta)(X-Y a\sin^2\theta)}
{\Sigma^2},~
F_{rt}=\frac{Q(r^2-a^2\cos^2\theta)}{\Sigma^2}, \nonumber \\
F_{t\theta}&=&\frac{Q ra\sin2\theta}{\Sigma^2},~
F_{\phi r}=\frac{Q a(r^2-a^2\cos^2\theta)\sin^2\theta}{\Sigma^2},~
F_{\theta\phi}=\frac{Q ar(r^2+a^2)\sin2\theta}{\Sigma^2}.
\end{eqnarray}
Since the Hamiltonian constraint (\ref{Hamiltonc}) is invariant
under (\ref{toBL}) with the transformed potential one-form
(\ref{potential:trans}), we only need to check the Maxwell
equation (\ref{fMaxwell}).
Indeed, the above potential one-form (\ref{potential:trans})
satisfies the Maxwell equation (\ref{fMaxwell}) with the metric
(\ref{BLT=0}) and (\ref{Hwithcharge}).
%


Now, we would like to examine the gravitational effect on the
brane due to the extra dimension.
To do that we will calculate the projected Weyl
tensor $E_{\mu\nu}$ in (\ref{grave}) using our potential one-form (\ref{potential:trans})
 and the metric (\ref{BLT=0}) with (\ref{Hwithcharge}).
In the charged rotating case the tensor $E_{\mu\nu}$ is quite complicated
to tell anything definite. For instance, $E^{r}_{~r}$ component is given by
\begin{eqnarray}
E^{r}_{~r}\!\!\!&=&\!\!\!-\frac{m_c^2}{\Sigma^2\Delta}
[~6a^2(~r^4(1-3\cos^2\theta)-2r^2a^2\cos^2\theta
\sin^2\theta+a^4\cos^4\theta(21-19\cos^2\theta)~)
\nonumber\\
\!\!\!&+&\!\!\!Q^2(r^4+6r^2a^2\cos^2\theta-3a^4\cos^4\theta)]
+\frac{Q^2}{\Sigma^2\Delta}(r^2-2Mr+Q^2).
\end{eqnarray}
In the non-rotating limit ($a \rightarrow 0$), with  (\ref{Hwithcharge}) and
(\ref{potential:trans}) the gravitational field equation (\ref{grave}) becomes
\begin{equation}
\label{ein}
\kappa_{4}^2 T_{\mu\nu}=-E_{\mu\nu},
\end{equation}
where $ T_{\mu\nu}$ is calculated to be the same energy-momentum tensor
as in the conventional four dimensional charged black hole case.
This tells us that there is no gravitational effect on the brane due to the extra dimension in the non-rotating charged case.
%


To examine the horizon structure of the metric given by (\ref{BLT=0}) and
(\ref{Hwithcharge}), we write the governing equation for the
radius of horizon
\begin{equation}
\label{horizon:conq}
\Delta=r^2+a^2+Q^2-2Mr-m_{c}^2(r^4+6r^2a^2\cos^{2}\theta-3a^4
\cos^{4}\theta)=0.
\end{equation}
The solution for the above equation provides three
horizons located at $r'_{\pm}$ and $r'_{CH}$ when the total
mass $M$ lies in the range $M_{1e}'|_{\theta=\pi/2} \leq M
\leq M_{2e}'|_{\theta=0,\pi}$ where $M_{1e}'$ and $M_{2e}'$
are given by
\begin{eqnarray}
\label{extrememass}
M_{1e}'=\frac{1}{3\sqrt{6}m_{c}} \sqrt{\alpha'-A'^{3/2}},~~
M_{2e}'=\frac{1}{3\sqrt{6}m_{c}} \sqrt{\alpha'+A'^{3/2}}.
\end{eqnarray}
Here,
\begin{equation}
\alpha'=1+36m_{c}^2 (a^2+Q^2)-18m_{c}^2 a^2 \cos^2
\theta(1+12m_{c}^2a^2+12m_c^2Q^2)+216m_{c}^4 a^4\cos^4
\theta(1-4m_{c}^2a^2\cos^2 \theta)
\end{equation}
and
\begin{equation}
A'=1-12m_{c}^2(a^2+Q^2+a^2\cos^2\theta ).
\end{equation}
The explicit expressions of the horizons are as follows:
\begin{equation}
\label{horizon:q}
r'_{\pm}=\frac{1}{2m_c}\left(D'^{1/2}_{-} \pm \sqrt{D'_{+}
-4Mm_{c}D'^{-1/2}_{-} }\right),~
r'_{CH}=\frac{1}{2m_c}\left(-D'^{1/2}_{-} + \sqrt{D'_{+}
+4Mm_{c}D'^{-1/2}_{-} }\right),
\end{equation}
where
\begin{equation}
D'_{\pm}=C \pm \frac{1}{3}
\left[ C+A' \left(\frac{2}{B'+\sqrt{B'^2-4A'^3}}\right)^{1/3}
+\left(\frac{2}{B'+\sqrt{B'^2-4'A^3}}\right)^{-1/3}\right],
\end{equation}
and $B'=2C^3+72m_c^2 C(a^2+Q^2+3m_c^2a^4\cos^4\theta)
-108m^2_c M^2$.

Note that the horizons $r'_{\pm}$ and $r'_{CH}$ always have a
real positive value if the total mass lies between the masses
$M_{1e}'|_{\theta=\pi/2}$ and $M_{2e}'|_{\theta=0,\pi}$.
In the limit $r_{c} \rightarrow \infty$, we get $M'_{1e} \rightarrow a$,
$M'_{2e} \rightarrow \infty$, $r'_{\pm} \rightarrow M \pm
\sqrt{M^2-a^2-Q^2}$, and $r'_{CH} \rightarrow \infty$ independent of the
angle $\theta$ as in the rotating case.

 The defining condition $g_{tt}=0$ for the ergosphere in this case
 is given by
\begin{equation}
\label{ergoQ}
r^2+a^2\cos^2\theta+Q^2-2Mr-m_c^2(r^4+6r^2a^2\cos^{2}\theta
-3a^4 \cos^{4}\theta)= 0.
\end{equation}
  Setting $r^{'\pm}_{E}$ and $r^{' CH}_{E}$ as
\begin{equation}
r_E^{'\pm}=\frac{1}{2m_c}\left(\tilde{D'}^{1/2}_{-} \pm
\sqrt{\tilde{D'}_{+}-4Mm_{c}\tilde{D'}^{-1/2}_{-} }\right),~
r_E^{'CH}=\frac{1}{2m_c}\left(-\tilde{D'}^{1/2}_{-}
+ \sqrt{\tilde{D'}_{+}+4Mm_{c}\tilde{D'}^{-1/2}_{-} }\right),
\end{equation}
where
\begin{equation}
\tilde{D'}_{\pm}=C \pm \frac{1}{3} \left[ C+\tilde{A'}
\left(\frac{2}{\tilde{B'}+\sqrt{\tilde{B'}^2-4\tilde{A'}^3}}
\right)^{1/3}+\left(\frac{2}{\tilde{B'}+\sqrt{\tilde{B'}^2
-4\tilde{A'}^3}}\right)^{-1/3}\right] ,
\end{equation}
with
\begin{eqnarray*}
 \tilde{A'} & = & 1-12m_c^2Q^2-24m_c^2 a^2\cos^2\theta , \\
\tilde{B'} & = & 2C^3+72m_c^2 C(Q^2+a^2\cos^2\theta+3m_c^2a^4
\cos^4\theta)-108m^2_c M^2,
\end{eqnarray*}
we get the same relation as in the non-charged
rotating case
\begin{equation}
\label{inequal} r_E^{'-} \leq r'_{-} < r'_{+} \leq r_E^{'+} <
r_E^{'CH} \leq r'_{CH},
\end{equation}
and the ergosphere lies in the region $r'_{+}< r< r_E^{'+}$ and
$r_E^{'CH} <r<r'_{CH}$  coinciding  with the horizon  at
$\theta=0,\pi$.

\section{Discussion}

In this paper we considered charged rotating black holes on a 3-brane in
the DGP model. Assuming a $Z_2$-symmetry across the brane and
with a stationary and axisymmetric metric ansatz on the brane,
we solved the constraint equations of (4+1)-dimensional gravity
to find a metric for charged rotating black
hole on the brane.

 First, we obtain a particular solution of the Kerr-Newman-de Sitter
type in the Kerr-Schild form, which
corresponds to the so-called accelerated branch of
the DGP model.

Then, in order to find the event horizon of the black hole, we introduce a
modified version of Boyer-Lindquist coordinates.
The Hamiltonian constraint equation
is quite complicated to solve, even compared with the RS model case \cite{ag:prd},
and not preserved under the conventional Boyer-Lindquist transformation.
Thus in order to use the obtained Kerr-Schild type solution, we have to
introduce a modified transformation which preserves the constraint equation.

In the case of the RS model, the authors of
\cite{ag:prd} devised a transformation for a given fixed angle $\theta$,
and showed that the equations are preserved under their transformation
thereby the metric function $H$ remains as a solution of the constraint equation.
However, with this type of transformation the coordinates patches for different $\theta$
angles belong to differently transformed coordinates, and it makes hard to view
the obtained event horizon in a single consistent picture.
In order to avoid this kind of problem, we use a slightly modified
version of Boyer-Lindquist coordinate transformation which covers
the entire $\theta$ angle while the solution obtained in the
Kerr-Schild form can still be used.
In this solution, the structure of the horizon is
very similar to that of the Kerr-Newman-de Sitter black hole in general
relativity except for the $\theta$-angle dependence.
When the crossover scale $r_c$ approaches infinity,
the $\theta$-angle dependence of the horizon disappears
and the solution reduces
to that of the Kerr-Newman black hole in general relativity.


 Finally, we discuss a possible bulk solution consistent with our on-brane solution.
 For this, here we limit ourselves to the non-rotating limit
 to make our discussion tractable.
 Rather than following the strategy of extending the on-brane solution to the bulk,
 we try directly to find a bulk solution consistent with our on-brane solution.
For the most simple uncharged case,
we find that the following 5D metric satisfies the 5D field equations, (\ref{eom:munu}):
\begin{eqnarray}
\label{bulkmet}
ds^2&=&e^{-2m_c|y|}(dy^2+h_{\mu\nu}dx^{\mu}dx^{\nu}),
\end{eqnarray}
where
\begin{eqnarray}
h_{\mu\nu}dx^{\mu}dx^{\nu}&=&
-\left(1-\frac{2Mr+m_c^2r^4}{r^2}\right)dt^2
+\left(1-\frac{2Mr+m_c^2r^4}{r^2}\right)^{-1}dr^2
\nonumber \\
&&+r^2 (d\theta^2+\sin^{2}\theta d\phi^2).
\end{eqnarray}
Namely, the above metric satisfies the 5D Einstein equation in the bulk, $^{(5)}G_{AB}=0$, as well as
 the on-brane field equation, (\ref{grave}).
The projected Weyl tensor $E_{\mu\nu}$ obtained from the 5D metric (\ref{bulkmet}) vanishes,
and this is consistent with the previously obtained relation (\ref{ein})
since the energy-momentum tensor vanishes in this case.
Therefore, in the uncharged case we can say that our on-brane solution is consistent
with the above given bulk solution.

 For the charged case, the electro-magnetic field
 vanishes off the brane (in the bulk) by the set-up. 
 So far we could not find a bulk solution which smoothly matches the metric on the brane
 with the metric off the brane while reflects the discontinuity of electro-magnetic field 
 at the boundary which is non-zero on the brane and suddenly vanishes off the brane.
We now leave this challenging problem of finding consistent bulk solutions for the charged and rotating 
case as an open project and welcome anyone who is interested in.  
\\


\section*{Acknowledgments}

The authors thank KIAS for hospitality during the time that this
work was done.
D.L. and E.C.-Y. were supported by the Korea Research Foundation
Grant funded by the Korean Government(MOEHRD), KRF-2006-312-C00498.
M.Y. was supported by the Science Research
Center Program of the Korea Science and Engineering Foundation through
the Center for Quantum Spacetime (CQUeST) of Sogang University with
grant number R11-2005-021.
\\


\end{document}